\documentclass[
11pt,%
tightenlines,%
twoside,%
onecolumn,%
nofloats,%
nobibnotes,%
nofootinbib,%
superscriptaddress,%
noshowpacs,%
centertags]%
{revtex4}
\usepackage{amssymb}
\usepackage{natbib}
\usepackage{amsmath}
\usepackage{amsfonts}
\usepackage{graphicx}
\usepackage{mathrsfs}
\usepackage{dcolumn}
\usepackage{bm}
\usepackage{color}
\usepackage{cancel}
\usepackage{mathbbol}
\usepackage{epsfig}
\usepackage{units}
\usepackage{esint}

\begin{document}

\title{Simulating quantum dynamics: \\
Evolution of algorithms in the HPC context}

\author{\firstname{I.}~\surname{Meyerov}}
\email[E-mail: ]{meerov@vmk.unn.ru} \affiliation{Department of
Software and Supercomputing Technologies, Lobachevsky University,
603950, Nizhny Novgorod, Russia}

\author{\firstname{A.}~\surname{Liniov}}
\email[E-mail: ]{alin@unn.ru} \affiliation{Department of Software
Engineering, Lobachevsky University, 603950, Nizhny Novgorod,
Russia}

\author{\firstname{M.}~\surname{Ivanchenko}}
\email[E-mail: ]{ivanchenko.mv@gmail.com} \affiliation{Department of
Applied Mathematics, Lobachevsky University, 603950, Nizhny
Novgorod, Russia}

\author{\firstname{S.}~\surname{Denisov}}
\email[E-mail: ]{sergiyde@oslomet.no} \affiliation{Department of
Computer Science, Oslo Metropolitan University, N-0130, Oslo,
Norway} 



\begin{abstract} 
Due to  complexity of the systems and processes it addresses, the
development  of computational quantum physics is influenced by the
progress in computing technology. Here we overview the evolution,
from the late 1980s to the current year 2020,  of the algorithms
used to simulate dynamics of quantum systems. We put the emphasis on
implementation aspects and computational resource  scaling with the
model size and propagation time. Our mini-review is based on a
literature survey and our experience in implementing different types
of algorithms. 
\end{abstract}



\maketitle


\section{Introduction}

The agenda of computational quantum physics (CQP) is to provide  researchers with
tools to model quantum systems
on computers. Since most of the problems in quantum mechanics cannot
be solved analytically, numerical methods were always in demand and
played an important role in the development of quantum mechanics. In
the period between the late 1990s and early 2010s, the activity on
the CQP field was boosted by several waves of advances in
experimental physics, such as the appearance  of quantum
optics of ultracold matter (marked by the creation of the
Bose--Einstein condensate in a lab \cite{BEC}) and fast progress in
superconducting microwave technologies (resulted in the creation of
the first generation of quantum computer prototypes \cite{qubit}). 
Almost instantly, CQP turned to be not only a branch of theoretical quantum physics that
assists the latter in gaining  new knowledge but also a toolbox of
methods to design new experiments and blueprint quantum devices.

The new status of CQP strengthened the ties between quantum physics and high-performance computing (HPC) 
and changed the character of the research activity on the field. Starting from the 2010s,
a familiarity with cutting-edge computing technologies and knowledge of how to use them to handle larger and more complex models
became important elements of the professional expertise. By now,
CQP represents a synergetic combination of quantum physics, applied mathematics, and HPC,
in which the last component is no less important than the first two.

In this paper, we overview the evolution \cite{evolution} of the algorithms used for digital simulations of the dynamics of quantum systems.
We, therefore, do not discuss different diagonalization, renormalization, and variational techniques
used to find ground-state or/and first excited states (unless the corresponding technique is a part of the discussed simulation algorithm).
We put the emphasis on such computational aspects as the resource scaling, cluster implementation, and parallelization, and try to address them
in the context of the HPC technology development. The evolution is illustrated (Fig.~1) by adapting the idea of the Gartner
Hype Cycle for Emerging Technologies \cite{Gartner}.

Our overview is partially based on the literature survey. The
analysis of the publications revealed that it is very seldom that
the information on the details of algorithm implementations and computational
resources is provided, even  in additional materials and appendices.
Therefore we supplement the consideration by describing  our experience in
implementing  different simulation algorithms  
on supercomputers ``Lobachevsky'' (at Lobachevsky State University
of Nizhny Novgorod) and ``Lomonosov 2'' \cite{lomonosov} (at Moscow
State University).

The remainder of the paper is organized as follows. In Section 2 we discuss the evolution of the algorithms
sketched in Figure 1.
In Section 3 we address the first two generations of algorithms
designed for implementations on classical computers.
In Section 4 we discuss algorithms developed for digital quantum simulation (i.e., for implementations on 
digital  quantum computers \cite{nielsen}). In Section 5 we
outline recent advances related to the Machine Learning approach to QCP.
Finally, in Section 6 we  summarize the consideration and  discuss  future perspectives and possible trends.

\begin{figure}[t]
\includegraphics[width=0.75\textwidth]{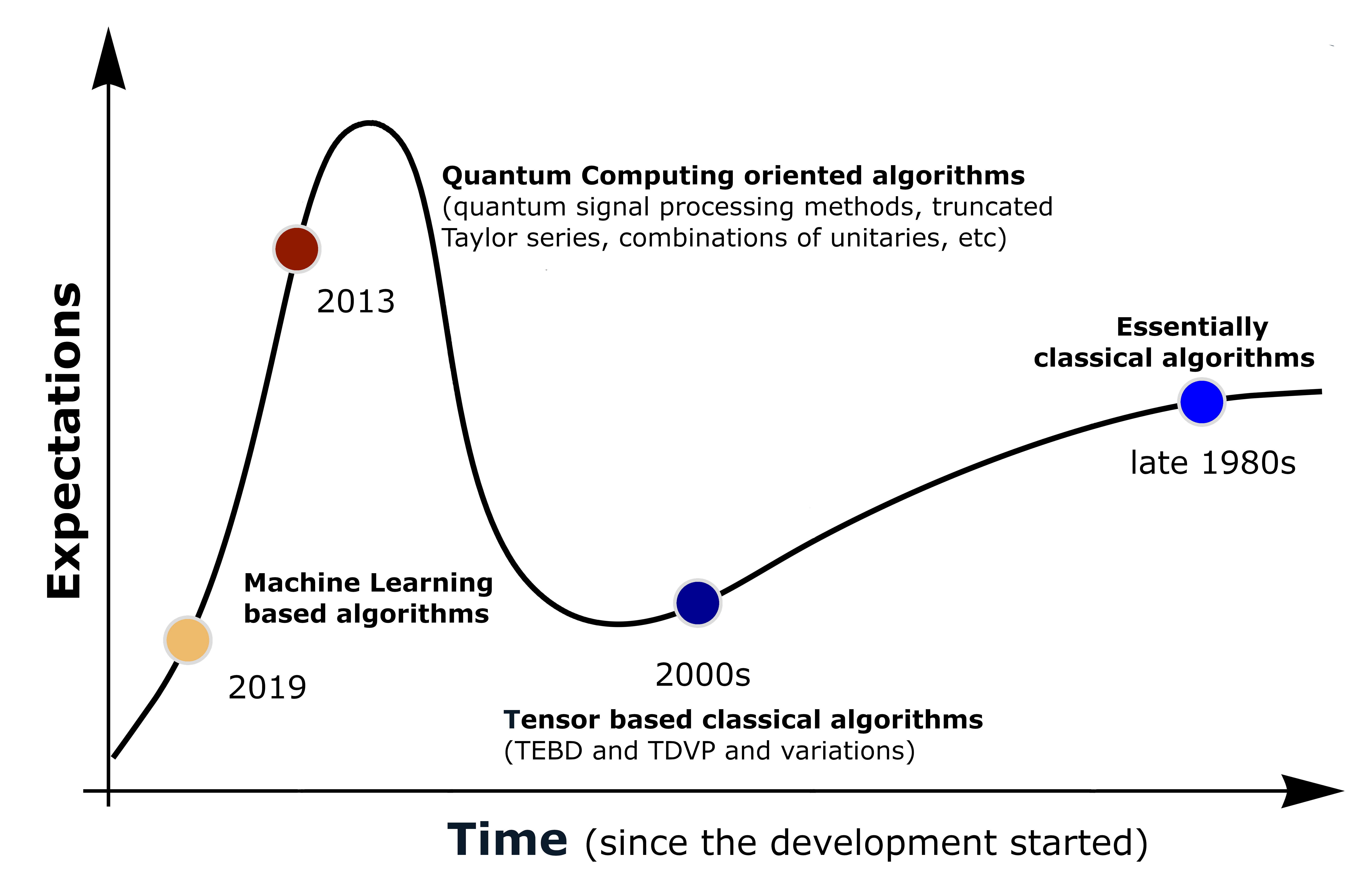}
\caption{Evolution curve of algorithms to simulate quantum dynamics.}\label{fig:1}
\end{figure}

\section{Evolution curve of simulation algorithms}

To model the dynamics of a quantum system, evolving in a Hilbert space $\mathcal{H}$  of dimension $dim~\mathcal{H} = N$, 
we have to integrate numerically
the initial value problem for one of the two  operator-valued differential equations. In the case of coherent unitary evolution,
it is the Schr\"{o}dinger equation,
\begin{eqnarray}
i\hbar |\dot{\varphi}(t)\rangle = \hat{H}(t)|\varphi(t)\rangle,
\end{eqnarray}
with the initial wave function $|\varphi(0)\rangle =
|\varphi_0\rangle$. The Hilbert space can be spanned with a basis,
$\{|\psi_j\rangle\},~j=1,2,...,N$. The expansion over the  basis
allows transforming the wave function into an $N$-dimensional
complex column vector, $|\varphi (t) \rangle = \sum
c_j(t)|\psi_j\rangle\ \Rightarrow [c_1(t),c_2(t),...,c_N(t)]^T$, and
the Hamiltonian operator $\hat{H}(t)$ into an $N$-dimensional
time-dependent Hermitian matrix. In the case of open evolution, when, e.g., 
the system is coupled to its environment, the  state of
the system is described with the density operator $\varrho(t)$. The evolution of this operator
is governed by a quantum Liouvillian  $\mathcal{L}(t)$
(most often of the so-called
Gorini--Kossakowski--Sudarshan--Lindblad (GKSL) form \cite{bookP}
but some other forms \cite{kohler} are also used),
\begin{eqnarray}
\dot{\varrho} (t)= \mathcal{L}(t)\varrho(t),
\end{eqnarray}
with the initial density operator $\varrho_0$. Often the density operator is also vectorized and transformed into $N^2$ (or $N^2-1$) complex vector
and the Liouvillian is then recast in the form of an $N^2$ (or $N^2-1$)-dimensional time-dependent matrix.

We adapt the idea of the Gartner Hype Cycle of Emerging Technologies \cite{Gartner} to illustrate the evolution of algorithms developed to integrate Eqs.~(1-2); see Fig.~(1).
The evolution consists of several stages and reflects the overall attitude of the CQP community towards an algorithm.
At first, a new promising algorithm is recognized as an ``Innovation Trigger''.
If there is growing confidence in  the community that the algorithm
has a potential and can be used to solve standing problems,  the activity and number of publications on the algorithm are going up
and this brings the latter to the ``Peak of Inflated Expectation''. The peak
is usually succeeded by the phase of sobering and realization of the algorithm limitations, and the community attitude slides down
to the ``Trough of Disillusionment''. Finally, if the algorithm survives this phase,
it enters  the ``Plateau of Productivity'' and becomes a part of the conventional CQP toolbox.
We address four different 'species', or (with a slight abuse of the terminology), \textit{generations} of the algorithms.

The first generation was remaining the only one until the late 1980s. At that time, the overwhelming majority of the considered models
describe single-particle systems  and an implementation of one of the standard schemes used to integrate
differential equations (though tailored to the symplectic structure of the Schr\"{o}dinger equation)
such as different variants of split-step, pseudo-spectral, and Krylov subspace methods \cite{propH}, was enough in most cases.
These algorithms remain an important  part of the  CQP toolbox and they are routinely used, e.g.,  in computational quantum chemistry.
During this initial stage of the evolution, computational aspects of algorithm implementations were not considered as important
and such questions as resource scalings and parallelization  were out of the focus.

The situation changed  with the rise of many-body  quantum physics in the late 1990s. This was the time when the
computational quantum community faced the ``Curse of Dimensionality''~\cite{curse}, which in this case turned to be an exponential
growth of the amount of data, needed to be stored,
and computational resources needed to process this data,  to simulate the dynamics of a many-body model,
with the number of 'bodies' (spins, qubits, photons, etc) the model is built of.

The appearance, at the beginning of the 2000s, of a new generation of algorithms \cite{tensor1,tensor2,tensor3},
based on the low-rank tensor approximation ideas \cite{oseledets},
was a substantial advance in the fight with the Curse.
The algorithms turned to be very successful when used to simulate many-body systems of linear topology characterized by a short-range entanglement
and  were a decisive factor in the development of new research fields such as many-body localization and transport in quantum disordered systems.
These algorithms [posses a substantial potential for parallelization and their appearance initiated the development 
of the HPC component in CQP. It has also
become clear to non-experts (though experts were aware of this limitation from the beginning)  that the propagation of a generic model with
unbounded growth of entanglement  is limited by a certain -- often  a very short -- time horizon  and so these
algorithms have a limited scope and should be used with care \cite{pitfall1,pitfall2}. On the evolution curve,
we place these algorithms (rather subjectively) to the ``Trough of Disillusionment'' region.

First two generations are \textit{classical}, i.e., the corresponding algorithms  were on purpose designed to simulate quantum models
on classical digital computers ~\cite{tensor1}.
There are two new CQP trends that are  catalyzed by recent technological advances.
First, it is the appearance of quantum computer prototypes  on the IT market and their accessibility to the QCP researchers  \cite{cloud}.
Even though the devices manufactured by Rigetti, IBM, Google, and Microsoft belongs to the generation of the so-called
'Noisy Intermediate-Scale Quantum (NISQ)' processors \cite{nisq}, their advent boosted the development of a  new generation of simulation algorithms
\cite{qp1,qp2,qp3} designed for implementations on future digital quantum computers. These algorithms go beyond the generic
Trotterization ideology \cite{simulator} and  can be used to simulate dynamics of quantum models of different types,
not only systems of qubits or spins.

Finally, the rise of Machine Learning (which currently is modifying the whole paradigm of computational physics \cite{ML})
brought new perspectives. Machine learning strategies and artificial neural networks (ANNs) were proposed as means
to compactify the descriptions of many-body states and recognize different quantum
phases \cite{ML}.
It was realized that Matrix Product States (MPS), which serve a foundation for the tensor-based simulation algorithms,
represent a particular class of the ANN states. There are several works appeared recently in which ANNs were suggested as the tool to 
simulate the dynamics of many-body models \cite{ANN1,heyl,guit}, even beyond the time scales that can be reached with the most advanced tensor-based
algorithms~\cite{heyl}.

\section{Classical algorithms}

\subsection{Performance optimization and parallelization: Why it is important?}

Modern computing systems are becoming more and more complex and heterogeneous. This requires
new methods for performance optimization and parallelization of scientific software.
Such methods are supposed to utilize resources of cluster systems with multi- and many-core CPUs at the following levels:

\begin{itemize}
    \item Computational nodes on distributed memory: reducing the overhead of data transfers, load balancing;
    \item Computational cores on shared memory: Non-uniform memory access (NUMA)-aware memory usage, cache optimizations, load balancing;
    \item Single instruction multiple data (SIMD) units in computational cores: code vectorization.
\end{itemize}

In this regard, we often need to develop a hybrid parallel algorithm by using MPI~$+$~OpenMP/TBB/\ldots
technologies  with efficient use of SIMD instructions.
When using graphic processing units (GPUs), we can potentially make computations
faster if the task is tidied up to a limited amount of GPU's internal memory but the code development often becomes more intricate.

When implementing already existing algorithms, the problems of optimizing performance and efficient parallelization  comes to the fore.
Digital simulations of quantum model by using an algorithm belonging to the first generation is computationally a very expensive task.
To simulate the dynamics of an open quantum system, described
by the  master equation, Eq.~(2), we have to operate with matrices of the size $N^2 \times N^2$.
This can be problematic already starting $N \sim 10^3$. It is often
necessary to repeat simulations many times, for example, in order to analyze the behavior of a model system at a
large number of points in the parameter space.

In this regard, there is a need to develop and use custom data structures and parallel algorithms
that allow solving the problem on a cluster in an acceptable
time and fit into the imposed memory limits.
To date, several software toolkits to simulate  dynamics of quantum systems on classical computers have been developed. For example, an
open source software QuTiP (Quantum Toolbox in Python) \cite{QuTiP} allows one to model dynamics of open quantum systems.
Being developed in Python, QuTiP is based on the high-performance libraries Numpy, Scipy, and Cython, which take care of performance.
There is also a TBTK package \cite{TBTK},  `` an open-source C++ framework for
modeling and solving problems formulated using the language of second quantization''.
Another software, WavePacket, is the MATLAB code, designed to simulate dynamics of coherent \cite{WP1} and open \cite{WP2} quantum models.
The versatility and variability of methods implemented in these and other software toolkits are an undoubted advantage;
however, achieving high efficiency of using parallel supercomputers requires taking into account the features of the specific task and the model used.
In this regard, we consider some examples demonstrating how the development of custom data structures and tailored parallel algorithms can lead to a
significant performance gain.

\subsection{Essentially classical algorithms}

When realizing some of our CQP projects, we designed and
implemented two types of algorithms belonging to the first generation,  one to simulate unitary dynamics of quantum models with an explicitly
time-dependent Hamiltonians \cite{Magnus1} and another one to simulate  dynamics of open models with  time-dependent
Liouvillians \cite{FBasis1,FBasis2}.

In \cite{Magnus1} we considered the problem of parallelization of an algorithm to
calculate eigenstates of large quantum models with Hamiltonians that are modulated periodically in time. 
This demanded numerical propagation of many initial states up to the time equal to the period of modulations.
The main part of the algorithm combines the Magnus expansion of the time-dependent system Hamiltonian with the Chebyshev expansion of an operator exponent \cite{Blanes}.
In this method, the computations are based on linear algebra operations, in particular, on the dense matrix-vector multiplications.
Parallelization of calculations on distributed memory is performed trivially with an ideal load balancing by dividing the vectors of
initial conditions among the processes
with subsequent propagation and collection of the results and their relatively simple processing on one master node.
The main optimization applicable
in the calculations at each node of the cluster is the simultaneous propagation of all vectors of initial conditions,
which allows switching from matrix-vector multiplications
to matrix-matrix multiplications.
This optimization speeds up calculations by an order of magnitude due to more efficient work with memory
hierarchy with the same number of floating-point operations. Vectorization
of computations is achieved through the use of high-performance BLAS implementations, for example,
from the Intel Math Kernel Library. Note that it is a common approach: the switching
to the third level BLAS and the use of highly-optimized mathematical software usually allows
achieving near optimal performance. In this problem, the code can be easily ported to
GPUs by switching to cuBLAS with a significant reduction in computation time.

Similar methods are considered in the paper \cite{Magnus2}, which explains how to efficiently implement the Magnus
integrator with Leja interpolation. The paper explores the commutator
free method, which replaces computationally intensive matrix multiplications with matrix-vector multiplications thus
reducing the number of floating point operations. The authors state
that such their method is better from the numerical analysis point of view, but it does not fit  the architecture
of modern parallel CPUs and GPUs, which eliminates its advantages.
The authors also demonstrated that GPUs can speed up calculations by an order of magnitude in a dense testbed problem.

Another example of the effective use of specific problem features is given  in  \cite{FBasis1} and \cite{FBasis2}.
The implemented algorithm transforms the Lindblad equation into a system of linear ordinary differential equations with
real coefficients by using the generalized Gell-Mann matrices as a basis \cite{FBasis1}. Such a system can be
propagated forward in time with one of the standard high-order integration methods. It was demonstrated \cite{FBasis1}, that a naive method of
computing this expansion requires enormous computation time and memory,
while the construction of the specific data structures and methods for their computing, based on counting
the only nonzero elements, can significantly reduce both computational complexity and memory usage. In this
method, sorting algorithms and dense \& sparse BLAS operations are the main mathematical kernels. Once again, we can
profit from high-performance mathematical software libraries. In \cite{FBasis2} it was shown  that these algorithms can also be
parallelized for cluster-based implementations. The main goal of parallelization in this case is not to reduce the computation
time, but to reduce the memory costs per each cluster node. As a result, it
was possible to simulate the model with $N=200$ states (dense case) and $N=2000$ states (sparse case) using 25 
nodes of a cluster with 64~GB RAM per node.

In this section, we would like also mention the Krylov subspace method \cite{krylov1,krylov2} which remains popular
in the CQP community. The method proved to be very efficient, from the point of memory size, when the average number of Krylov vectors on every
propagation step is much smaller than $N$. The corresponding algorithms are part of an open-source package
\cite{MPS} and the results of its performance analysis can be found in \cite{jaschke}.
A parallel supercomputer implementation was reported in \cite{brenes}.

\subsection{Tensor-based classical simulation algorithms}

The total length $L$ of the description (the number of complex coefficients required to specify a quantum state) of a model system
consisting of  $M$ elements,
each one with $d$ degrees of freedom,  scales as $L(M) \sim N = d^{M}$. To specify an \textit{arbitrary} state of a
system of $50$ qubits, we need $2^{50} \approx 10^{15}$ complex numbers. With the double-precision format,
this exceeds the memory capacity of the supercomputer ``Lomonosov 2''~\cite{lomonosov}.
For an open quantum model, the complexity squares: to specify a density
operator we need $L(M) \sim d^{2M}$ real-valued parameters.

Singular Value Decomposition (SVD) is one of the best tools to reduce the amount of stored data  when
dealing with large matrices \cite{samet}. Its generalization to tensors,
so-called Tensor-Train (TT) decomposition \cite{oseledets}, also turned to be very effective when dealing with tensors.
In the physical literature, this decomposition is commonly referred
to as Matrix Product State (MPS; in the case of pure states) or Matrix Product Operator (MPO, in the case of mixed states) representation \cite{schollw}.
While two names are used simultaneously (though in different fields), the underlying mathematical structure is basically the same \cite{tensor3}.
The MPS/MPO/TT approach allows us to reduce the growth of the description of \textit{some} many-body states
to a linear scaling $L(M) \sim M$ \cite{oseledets}.

\begin{figure}[t]
    \includegraphics[width=0.7\textwidth]{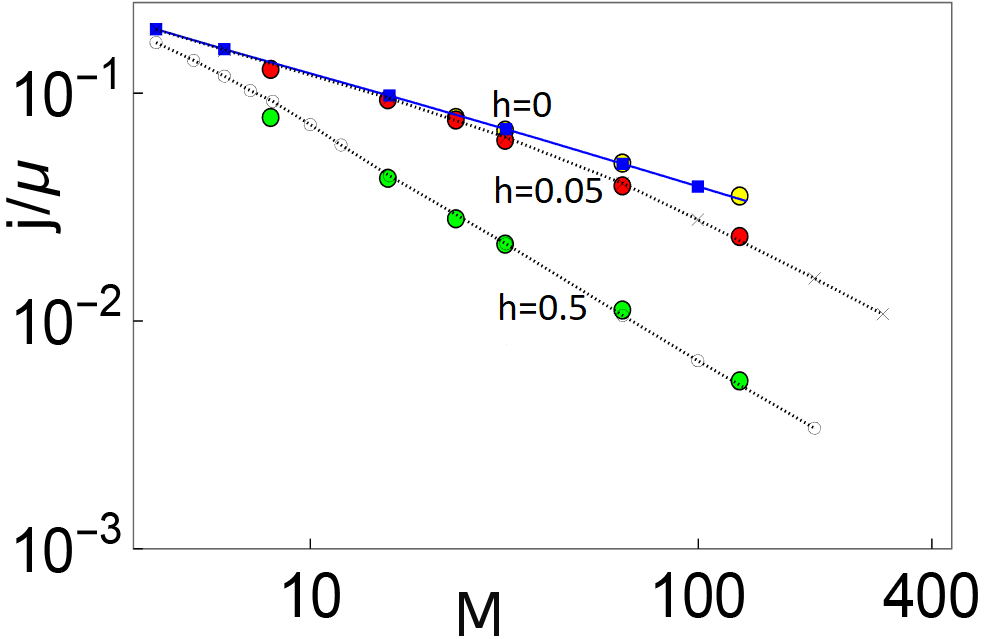}
\caption{Scaling of the spin current through a disordered spin 
    chain with $M$ spins for different values of disorder strength $h$.
        Our results \cite{TT2} (big filled circles) are plotted on top of the results reported in ~\cite{TT1}.}\label{fig:1}
\end{figure}

The MPS/MPO representation can also be used for an effective propagation of quantum many-body models. Following
the Time-Evolving Block Decimation (TEBD)
method \cite{tensor1}, the description of the state, obtained after every propagation step,
is reduced to a fixed  length $L_{\mathrm{cut}}$.
The accuracy of the propagation is therefore controlled by the value of  $L_{\mathrm{cut}}$: If the information is thrown
out after the reduction is substantial, the propagation is bad and leads to a wrong result. Otherwise, it is good.
Some many-body systems 'behave' well during the TEBD propagation and so the amount of neglected information
is tolerable  (we are not going to discuss physical properties underlying such a 'good behavior' and
refer the reader to the extensive literature on the subject; see. e.g., Ref.~\cite{schollw}).

In our recent work \cite{TT2},  we tried to reproduce the results reported in ~\cite{TT1},
where a disordered chain of $M$ spins, with a next-neighbor coupling and two ``thermal reservoirs'',
acting on the two end spins, was used as the model.
The transport of the spin charge through the chain in the stationary regime was considered and
the spin current scaling with the chain length $M$ was analyzed.
The results for $M=400$ were reported; see Fig.~2.  This is an unprecedented  size for a many-body open quantum model (to the best of our knowledge at the time).
The complexity of the computational experiments was increased by the fact that the model had to be propagated over a long
time in order to reach the stationary state.  Finally, to obtain scaling dependencies, an averaging over many disorder realizations  was performed.
At the same time, the work reports no details of numerical
simulations (even the value of such an important parameter as the bond dimension $R$ was not specified).

We  implemented a parallel version of the TEBD algorithm by using
the MPI technology and the standard master-worker scheme. The
computational experiments have been performed on the
``Lobachevsky'' cluster, with a $2 \times8$-core Intel Xeon CPU
E5-2660, 2.20~GHz, 64~GB RAM, Infiniband QDR interconnect. The code
was compiled with the Intel C++ Compiler, Intel Math Kernel Library
and Intel MPI from the Intel Parallel Studio XE suite of development
tools and the Armadillo library. We were able to reach reproduce the
results for $M=128$ spins (see Fig. 2),  by running our code on four
nodes of the cluster (one MPI-process per CPU core, 64 MPI processes
overall). Total computation time for a single disorder realization
was $143~s$.

The Suzuki--Trotter decomposition \cite{simulator} is one of the key
ingredients of the TEBD scheme. In \cite{TS1}, the authors
investigated the use of CPUs and GPUs to optimize the performance of
one round of the decomposition. It was demonstrated that it is
possible to obtain a significant performance gain through code
vectorization and optimization of memory usage patterns. By  using
GPU calculations in some model problems it was possible to reduce
the computation time by one order of magnitude. As a subsequent
development, a massively parallel version of the solver was
presented in ~\cite{TS2} and its functionality was further
expanded in ~\cite{TS3}.

There is a family of more advanced algorithms realizing the so-called Time-Dependent Variational Principle (TDVP) \cite{tensor2,tensor3}.
The main difference from the TEBD family is that the model evolution is confined to the subspace of the MPS (MPO) states and
therefore there is no need to perform the truncation after every steep (or a fixed number of steps).
This does not mean, however, that the TDVP propagation is numerically exact. In fact,
a model system could leave the MPS/MP subspace when evolving in its  Hilbert space; in other words,
the confinement is imposed not by the physics of the model  but by the algorithm.
Currently, the TDVP-based algorithms are considered to be the most advanced simulation algorithms in the CQP
community working with many-body  models.

In \cite{shol1}, a detailed  analysis of the single-node performance of TEBD and  TDVP implementations
is presented. Both implementations were tested on ''a single core of a Xeon E5-2630 v4 with 64 GB of RAM``
and so the parallelization potential was not addressed. In  a very recent work ~\cite{jak}, a cluster implementation of
the TDVP scheme to simulate spin chain models with long-range interactions was presented. It is reported that
the code ''scales well up to 32 processes, with parallel efficiencies as high as 86 percent`` and results of simulations 
for 201-site Heisenberg $XXX$ spin chain with quadratically decaying interactions are presented.

Finally, there is a generalization of the MPS concept to higher-dimensional models, the so-called Projected Entangled Pair States
(PEPS) \cite{murg}. The thermodynamic limit is addressed with infinite-PEPS (iPEPS) \cite{vidal2}
and there are several algorithms implementing the iPEPS representation to simulate the dynamics of infinite-size lattice models.
Here we mention a relatively recent  work, in which some computational aspects of iPEPS-based simulations
are discussed \cite{iPEPS}. However, no information on the computation resources used for simulations is provided.
We guess that the simulations were performed on a single core.

\section{Algorithms for digital quantum simulations}

The qubit architecture of the quantum processors manufactured by  Rigetti, IBM, Google,
and Microsoft, gives hope that Feynman's idea \cite{feynman} of modeling  quantum systems on quantum computers will be realized in near future.
However, future full-fledged digital quantum computers require new algorithms whose logics is principally different from
the logics of algorithms currently used in computational quantum physics -- simply
because the latter are designed to be implemented on classical (super)computers.

There  is a new QC-oriented paradigm that already brought several algorithms to simulate the
dynamics of quantum systems, of different nature and genesis, on quantum processors of the qubit architecture; see, e.g.,
Refs.~\cite{qp2,qp3}. The corresponding research activity,  as well as the expectations placed on these algorithms, 
are going up steadily, being heated by the fast progress on the front of QC-technologies  and strong competition between 
the main players on the QC market.

Similar to their classical predecessors, quantum simulation algorithms can be characterized by scalings ''number of operations (gates)
vs time of propagation and/or size of the model and/or accuracy``. The first step in this
direction was reported in a recent work \cite{pnas}. A method to 'compile',  i.e., to minimize the number of gates, was tested with
several existing algorithms and a  one-dimensional nearest-neighbor   Heisenberg system was used as the   model.
The scaling ''number of gates vs the size of the model (number of spins)`` was addressed. It is not clear though
whether actual \textit{simulations} (e.g., by using a classical emulator) were performed.

There is  a need for more  quantitative results and evaluation tests of the algorithm performances on the intermediate scale, 30 -- 60 qubits,
ideal (i.e., noise-free) quantum processors \cite{mitigation}. Simulations of different models, ranging from textbook examples to technology-relevant
systems, could bridge quantum physics and quantum computing in a new way, by allowing, for example, to describe properties
of quantum models in terms of their \textit{quantum} computational scalability. Presently, such a research program  can only
be realized on classical computers. This immediately leads us to the question of the efficiency of cluster-based
emulators.

Modern supercomputers allow simulating quantum systems consisting of 38 \cite{quest2},
42 \cite{ref7}, 45 \cite{ref12}, 49 \cite{ref13}, possible 53 \cite{ref14}, and in special cases up to 64 qubits \cite{ref15}.
There is a variety of open-source packages \cite{ref1} that can be used to emulate quantum circuits  \cite{ref2,ref3,ref4,ref5},
execute quantum algorithms \cite{quest2,ref7,ref8,ref9}, and even support a complete development
cycle from describing an algorithm to mapping it onto  specific quantum computer architecture, including optimization, verification,
and performance evaluation \cite{ref10}. Due to the various  characteristics \cite{ref11}, there is
an option to choose an emulator  most optimal  for  a specific task and available computing resources. We choose  QuEST \cite{quest1}, a recently
released ''open source, hybrid multi-threaded and distributed, GPU accelerated simulator of universal quantum circuits`` \cite{quest2}.
Even though the results of the performance analysis of QuEST are reported in Ref.~\cite{quest2}, we decided to do it independently.

Our experiments using the QuEST package show that simulations of
circuits consisting  of 30--36 qubits are possible on a single node,
with an evident limiting factor that is the amount of memory. For 30
qubits, about 18~GB is required, for 36 -- about 1.1~TB. The
distributed version allowed us to use the computing power and memory
of several nodes, for example,  we  were able to simulate (with a
speed-up factor $~20 - 30$ due to the MPI/Open MP use) up to 38 and
40 qubits on the supercomputers ''Lobachevsky'' and
``Lomonosov-2'', respectively. In our opinion, it is enough to
emulate digital quantum simulations with different scalable models
of non-qubit(spin) nature.

\section{Artificial neural network simulation algorithms}

There is an emerging trend in the CQP field initiated by the rise of the Machine Learning technologies \cite{ML}.
Among many other interesting findings, it was found that artificial neural networks (ANNs) can be
used to store highly entangled states of many-body systems  \cite{ML,ann1}.
It was also found that  many-body states which belong to the  MPS/MPO  (one-dimensional lattices) or
PEPS (two-dimensional lattices) classes represent sub-sets of the ANN states \cite{ann1, ann2}.

\begin{figure}[t]
\includegraphics[width=0.7\textwidth]{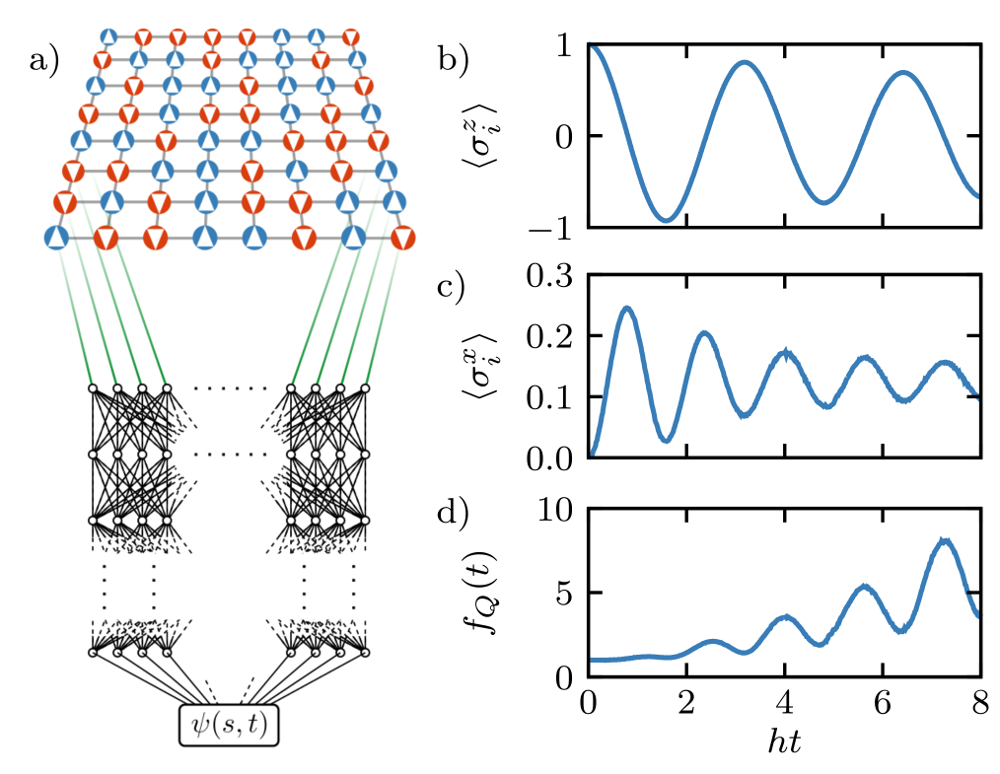}
\caption{ (a) Sketch of the propagation of a many-body model, the transverse-field square-lattice Ising system,
whose state is encoded  by a  deep convolution neural network. (b - c) Oscillations of the ferromagnetic order induced by the quench of the magnetic field at $t=0$.
(d) The so-called quantum Fisher information reveals the development of the multipartite entanglement in the system in the course of the propagation.
Courtesy of Markus Heyl.}\label{fig:1}
\end{figure}

There is an immediate question: Can we make the next step and use a trained ANN to propagate the state of a quantum model forward in time?
This idea was put forward and tested in the work by Carleo and Troer \cite{ANN1}.
They used a popular model, a square-lattice transversed-field Ising system,
and a special type of ANN, the so-called Restricted Boltzmann Machine \cite{MLtextbook}.
By quenching the magnetic field, a strongly non-equilibrium evolution was initiated. It was demonstrated that a trained
Reduced Boltzmann Machine is able to propagate the model with high accuracy (so that, in terms of observable the difference
from the numerically exact unitary propagation remained small) 
up to a time sufficient to see the quench-induced  oscillations in the ferromagnetic order.

In a recent work \cite{heyl},  a deep convolutional neural network (a broader class than  Restricted Boltzmann Machines) was used.
The authors demonstrated that with it is possible to simulate the dynamics of the $2d$ transversed-field Ising model
with higher accuracy and for a time longer
than can be achieved with a tensor-based algorithm \cite{iPEPS} (presumably, by using the same computation resources).
The simulations were performed on an Intel Xeon E5-2680 server by using the standard MPI/OpenMP technology. An alternative implementation on
a GPU VIDIA V100 was tested and turned to be ``30 times faster than our OpenMP implementation on 20 cores and 
still 15 times faster than an ideal parallelization on 20 CPU cores''
\cite{heyl}.  Similar results,  by using the same model (but of a smaller size) 
and  a slightly different ANN approach, were also reported recently in~\cite{guit}.

\section{Conclusion}

We tried to  overview the evolution of the numerical algorithms developed to simulate dynamics of quantum systems and  discussed this evolution
in the context of the HPC development. Evidently, there are algorithms that remained outside the classification 
presented with Fig.~1 and which therefore  were not addressed.

For example, there is  the so-called ''quantum trajectories`` (QT) or ''quantum jump``
method (also known as the ''Monte Carlo wave function`` method) \cite{bookP},
which allows transforming  the  numerical  integration of the master equation,
Eq.~ (2), into a task of statistical sampling over an ensemble of quantum trajectories. Thus, we could  deal with an $N$-dimensional vector
instead of an $N$-dimensional matrix. The price to be paid for this reduction is that we have to sample over many realizations.
However, the sampling is an embarrassingly parallel problem and thus we can  benefit substantially from  the  use  of  a cluster.
In Refs. \cite{qt1,qt2} we consider the problem of reaching the
asymptotic state of a non-equilibrium open quantum model by using the QT-algorithm.
We demonstrated  that on a comparatively small cluster it is possible to propagate  models  with $N = 2000$ states. The
parallelization for both shared and distributed memory is straightforward, which is typical for Monte Carlo
methods. The main difficulty there lies in the efficient use of the memory hierarchy, since in a straightforward
implementation, not very effective matrix-vector multiplications are the dominant operations. It was shown in Ref.~\cite{qt2}
that by using a specially developed algorithm, it was possible to group many matrix-vector multiplications into
equivalent matrix multiplications, which led to  a $17$-fold acceleration.

As respect to the third generation,  the question ''What the term 'parallelization' could mean
in this case?'' is of interest. Are there simulation algorithms which can be effectively implemented on several digital
quantum processors, that are wired in a \textit{classical} way? (it might be that this question was already addressed in the
literature and we are simply ignorant of this fact). Another interesting direction is the
development of 'quantum compilers' and 'quantum software' \cite{troyer,coles}.
We hope that the researchers working on quantum software will soon address parallelization aspects.

Finally, we would like to discuss possible future trends related to the last, fourth generation.
Even though it is hard to gauge the potential of the ML paradigm in the CQP context,
the first results look very promising and inspiring.
At this point the questions on how to use supercomputers for training ANNs and the subsequent high-performance inference become relevant.
Currently, systems based on GPUs are mainly to train ANNs. The GPU architecture is optimal for a large number of the same
computationally intensive operations on different data sets that occur during training. At the same time, for ANN high-performance inference, we can effectively use
also CPUs, whose instruction set have recently extended by the Intel Deep Learning Boost instructions including Vector Neural Network Instructions (VNNI)
that enable INT8 deep learning inference support \cite{22}. Along with traditional architectures, new technological developments  are constantly emerging.
The initiative of Intel Corporation, which plans to release a new GPU optimized for AI and high-performance computing \cite{23}, and
the development of the Graphcore company, which released the Graphcore IPU (Intelligence Processing Unit) designed for AI algorithms \cite{24} seem to be very relevant in this respect.
We are confident that during the next decade we will witness the rapid development of hardware and software tools focusing on both 
high-performance scientific computing
and the use of Machine Learning technologies.  
Hopefully, this progress will also affect the research activity in the QCP field.

\begin{acknowledgments}

The work is supported by the Russian Science Foundation via Grant
No. 19-72-20086. The research is carried out using the equipment of
the shared research facilities of HPC computing resources at
Lomonosov Moscow State University \cite{lomonosov} and the
Lobachevsky supercomputer at Lobachevsky University of Nizhny Novgorod.

\end{acknowledgments}

\end{document}